\documentclass[aps,prl,reprint,showpacs,floatfix,amsmath,amssymb,
superscriptaddress,showkeys,groupedaddress]{revtex4-1}

\usepackage{graphicx}
\usepackage{epstopdf}
\usepackage{color}
\usepackage{hyperref}
\usepackage{bm}
\usepackage{printlen}
\usepackage{units}
\usepackage{bbm}

\renewcommand{\vec}[1]{\mathbf{#1}}

\newcommand{\no}{\noindent}

\begin{document}

\title{Origin of Magic Angles in Twisted Bilayer Graphene}

\author{Grigory Tarnopolsky}    

\affiliation{Department of Physics, Harvard University, Cambridge, MA 02138}

\author{Alex J.  Kruchkov}
\affiliation{Department of Physics, Harvard University, Cambridge, MA 02138}

\author{Ashvin Vishwanath}
\affiliation{Department of Physics, Harvard University, Cambridge, MA 02138}

\date{\today}

\begin{abstract}
Twisted Bilayer graphene (TBG) is known to feature isolated and relatively flat bands near charge neutrality, when tuned to special magic angles. However, different criteria for the magic angle such as the vanishing of Dirac speed, minimal bandwidth or maximal band gap to higher bands typically give different  results.  
Here we study a modified continuum model for TBG which has an infinite sequence of magic angles $\theta$ at which, we simultaneously find that  (i) the Dirac speed vanishes (ii) absolutely flat bands appear at neutrality and (iii) bandgaps to the excited bands are maximized. When parameterized in terms of $\alpha\sim 1/\theta, $  they recur with the simple periodicity of $\Delta \alpha \simeq 3/2$, which, beyond the first magic angle, differs from earlier calculations. Further, in this model we prove that the vanishing of the Dirac velocity ensures the exact flatness of the band and show that the flat band wave functions  are related to doubly-periodic functions composed of ratios of theta functions. 
 Also,  using perturbation theory up to $\alpha^8$ we capture important features of the first magic angle  $\theta\approx1.09^{\circ}$ ($\alpha \approx 0.586$), which precisely explains the numerical results. Finally, based on our model we discuss the prospects for observing the second magic angle in TBG.

\end{abstract}

\maketitle

\textit{Introduction.}  The recent discovery of correlated insulators and seemingly unconventional superconductivity in twisted bilayer graphene (TBG) \cite{Cao2018a,Cao2018b}
has revived interest in TBG \cite{Po2018,Thomson2018,Zou2018,Guinea2018a,Guinea2018,Carr2018,Su2018,Gonzalez2018,Wu2018b,Efimkin2018,Peng2018,Yuan2018,Xu2018,Riccardi2018,Ochi2018,Pal2018,Wu2018a,Zhang2018,Wu2018,Kang2018,Pizarro2018,Koshino2018,Kennes2018,Isobe2018,Rademaker2018,Qiao2018,Zarenia2018,Chung2018, Fidrysiak2018, Peltonen2018, Cazeaux2018, Wang2018}. 
Importantly,  these phenomena are observed in a narrow range of twist angles near $1.05^{\circ}$, i.e. the first \textit{magic angle}  where isolated and relatively flat band appear near the chemical potential \cite{Bistritzer2010, Bistritzer2011,Castro-Neto2012, Shallcross2010}. To date, the origin and recurrence of the magic angles is not clear, even whether it is a fundamental feature or the outcome of a numerical coincidence. By slightly modifying the continuum model of TBG we show that the appearance of the exactly flat band is a fundamental feature, with a remarkable mathematical structure that is exposed in this Letter.

\begin{figure}[b]
\includegraphics[width = 1.0 \columnwidth ]{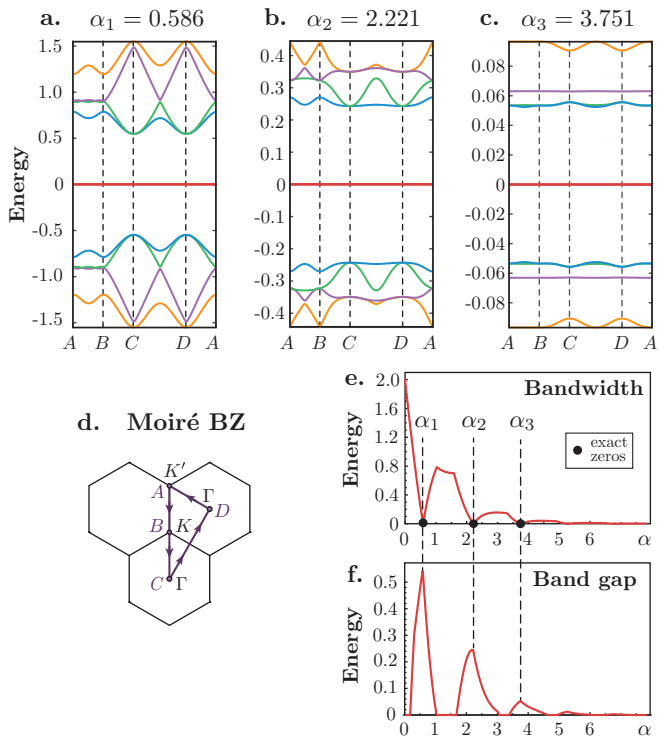}
\caption{Absolutely flat bands in  continuum TBG Hamiltonian (\ref{contmod}) with $w_{0}=0$ : in this model, the absolutely flat band appears at exact values of magic angles  $\alpha =0.586, 2.221,  3.751$, etc, where $\alpha= w_{1}/(v_{0}k_{\theta})$. Energy is given in dimensionless units   $\varepsilon=\alpha (E/w_{1})$. On subfigures (a-c), the band is numerically flat up to accuracy $10^{-16}$.  (d) Moir{\'e} Brillouin Zone. (e-f): The bandwidth for the lowest two bands vs $\alpha$.  The band width drops to exact zero at the set of magic angles. At the same points, we observe the maxima of the band gap.   }
\label{flatbands}
\end{figure}

Two graphene sheets placed on top of each other after twisting by a small relative angle form a long-period density wave pattern (Moir{\'e} pattern). For small angles $\theta$, the distinction between commensurate and incommensurate structures can be ignored, giving the lowest branch of Moir\'{e} periods as $L(\theta) = a_{0}/2 \sin \nicefrac{\theta}{2}$ ($a_{0}$ is  lattice constant). The question of the electronic spectrum at arbitrary small angles was studied by Lopes dos Santos et al.  \cite{Santos2007}, and by Bistritzer and MacDonald and others \cite{Bistritzer2010, Bistritzer2011, Castro-Neto2012, Shallcross2010} using a continuum model who reported appearance of the flat band at a set of magic angles starting at $\theta^{*}_1 \approx 1.05^{\circ}$.  More recently, several groups have used the continuum model and studied the role of topology of flat bands near the magic angles \cite{Po2018, Yuan2018, Kang2018, Koshino2018, Carr2018, Zou2018, Bernevig2018,Balents2018, Po2018a}. However, despite recent advances the origin of the magic angles in TBG remains mysterious.  

In this Letter, we refine the notion of the magic angle as the angle where the full band becomes flat; a more stringent requirement than the previous definition based on the vanishing of Fermi velocities at the Moir\'{e} Dirac points. For this, we introduce a slight variant of the traditionally studied continuum model, which is parameterized by two constants $w_0$ and $w_1$ which denote the inter-layer coupling in the $AA$ and $AB \& BA$ regions respectively. Early studies  \cite{Santos2007, Shallcross2010, Bistritzer2010, Bistritzer2011} set $w_0=w_1$  which we denote as the Bistritzer-MacDonald continuum model (BM-CM). Here instead we consider the continuum model with $w_0=0$, which acquires a chiral symmetry (a unitary particle-hole symmetry) which we call the {\em chirally symmetric continuum model} (CS-CM). Previously, the $w_0$ and $w_1$ terms were related to the scalar and non-Abelian vector potential perturbations of the Dirac fermions in Ref. \cite{Guinea2012} and our approximation corresponds to switching off the scalar part.  
In the CS-CM where we switch off $AA$ coupling completely but keep $AB$ and $BA$ finite, to our surprise a number of physical  phenomena converge to the magic angles. Not only do the Fermi velocities of the Moir\'{e} Dirac points vanish, but under these conditions the band becomes \textit{perfectly flat} at the recurrent set of magic angles (see Fig.~\ref{flatbands}). Moreover, the band gap to the excited bands is maximized at the same set of angles.  We conclude that this is the fundamental model that captures the essence of the magic angle phenomena, which continues to influence the physics on tuning away from this special point.  
Here we explore the CS-CM model in more detail, both analytically and numerically, and finally discuss restoring the $w_0$ perturbation.  While this model is motivated by its remarkable properties, interestingly we note  that in the presence of lattice relaxation, the $AB \& BA$ regions expand at the expense of the $AA$ regions effectively leading to a  continuum model for TBG with reduced  $AA$ coupling\cite{Koshino2018}, i.e. with $w_0/w_1<1$,  enhancing the direct relevance of our model to the experimentally realized system.

Yet another connection that we find is the close relation between the Bloch wave function at the Dirac points and the magic angles. At the magic angles, these spinor wave functions vanish at the $AB$ or $BA$ sites of the Moir\'{e} unit cell. Furthermore, the wave functions of the flatband at all crystal momenta can be analytically derived, if it is known at one point in the Brillouin Zone.  An interesting mathematical aspect here is that the wave function ratios are constructed from meromorphic doubly-periodic functions that are ratios of theta functions, similar to those appearing in the Quantum Hall Effect on the torus \cite{PhysRevB.31.2529}. 
 The CS-CM has a single coupling constant $\alpha = w_{1}/(2v_{0}k_D \sin(\theta/2))$ where $v_0$ and $k_D$ are the bare velocity and crystal momentum of graphene's Dirac fermions. We show that perturbation theory to high orders (up to $\alpha^8$)  matches with numerical results very accurately near the first magic angle. The sequence of magic angles that we find $\alpha = 0.586,\,2.221,\,3.751,\, 5.276, \,6.795, \, 8.313, \, 9.829,\, 11.345$,... reveals a remarkable asymptotic quasi-periodicity of $\Delta \alpha \simeq 3/2$. Comparing with the reported magic angles for the BM-CM, we see significant differences except for the first magic angle, see Table \ref{table:alpha}. We finally turn on the $AA$-coupling  and study numerically  how the bandwidth and gap evolves and discuss the possibility of studying the second magic angle in experiments.  
\begin{table}
 \caption{Comparison of magic angles in continuum model with $w_0=0$ (first row) and with $w_0=w_1$ (second row). Only the first magic angles correspond.}
 \label{table:alpha}
\begin{center}
 \begin{tabular}{c | c| c| c |c | c} 
 \hline
  \hline
  & \; $ \alpha_{1}$ \;  &\; $\alpha_{2}$ \; & \;  $\alpha_{3}$ \;  & \;  $\alpha_{4}$ \;  & \;  $\alpha_{5}$ \; \\ [0.5ex] 
 \hline
 CS-CM (here) & 0.586 & 2.221 & 3.75 & 5.28 & 6.80  \\ 
 \hline
  BM-CM (Ref. \cite{Bistritzer2010}) & 0.606 & 1.27 & 1.82 & 2.65 & 3.18 \\
 \hline
\end{tabular}
\end{center}
\end{table}

\textit{Continuum model  for Twisted Bilayer Graphene.} 
The  continuum model describing a single valley of TBG considers two layers of graphene described by an effective Dirac fields near $K,K'$ points of the moire (mini) Brillouin Zone, each rotated by an angle $\pm {\theta}/{2}$, and coupled through a Moir\'{e} potential $T(\vec r)$ \cite{Santos2007, Bistritzer2010, Po2018, Yuan2018}:
\begin{align}
H = \left(  \begin{array}{cc}
    -iv_{0} \bm{\sigma}_{\theta/2}\vec{\nabla} & T(\vec{r}) \\ 
     T^{\dag}(\vec{r}) & -iv_{0} \bm{\sigma}_{-\theta/2}\vec{\nabla}  \\ 
  \end{array}\right)\,, \label{contmod}
\end{align}
where $\bm{\sigma}_{\theta/2} = e^{-\frac{i\theta}{4}\sigma_{z}}(\sigma_{x},\sigma_{y})e^{\frac{i\theta}{4}\sigma_{z}}$,  $\nabla = (\partial_{x},\partial_{y})$  and 
\begin{align}
T(\vec{r}) =  \sum_{j=1}^{3}T_{j}e^{-i\vec{q}_{j}\vec{r}}
\end{align}
with $\vec{q}_{1}=k_{\theta}(0,-1)$, $\vec{q}_{2,3}=k_{\theta}(\pm\sqrt{3}/2,1/2)$ and 
\begin{align}
T_{j+1} = w_{0}\sigma_{0}+w_{1}\big(\cos(\phi j )\sigma_{x}+ \sin(\phi j )\sigma_{y}\big)\,, \label{Tj}
\end{align}
where $\phi=2\pi/3$, $k_{\theta}=2k_{D}\sin(\theta/2)$ is the Moir\'{e} modulation vector  and $k_{D}=4\pi/(3a_{0})$ is the magnitude of the Dirac wave vector, where $a_{0}$ is the lattice constant of monolayer graphene. The Hamiltonian (\ref{contmod}) 
acts on the spinor $\Phi(\vec{r}) = (\psi_{1},\chi_{1},\psi_{2},\chi_{2})^{\textrm{T}}$  and the indices $1,2$ represent the graphene layer. Here $w_0$ is responsible for the $AA$ coupling and $w_{1}$ is for $AB$ and $AB$ couplings.

\textit{ Chirally symmetric continuum model.}
In this Letter, we study a model obtained from Hamiltonian (\ref{contmod})  by setting $w_{0}=0$ \footnote{A different anti-unitary particle hole symmetry is realized by neglecting $\pm \theta/2$ rotations 
in the kinetic terms $\vec{\sigma}_{\pm \theta/2}\to \vec{\sigma}$, \cite{Balents2018, Bernevig2018} which is not considered here.}. Note, one can rotate the spinors to eliminate the rotation in the kinetic terms $\bm{\sigma}_{\pm \theta/2}\to \bm{\sigma}$ in the absence of the $w_0$ term. The dimensionless Hamiltonian now acts on a spinor 
$\Phi(\vec{r})=(\psi_{1},\psi_{2},\chi_{1},\chi_{2})^{\textrm{T}}$, 
and can be compactly written in the form

\begin{align}
\mathcal{H}= \left(  \begin{array}{cc}
    0& \mathcal{D}^{*}(-\vec{r}) \\ 
    \mathcal{D}(\vec{r}) &0 \\ 
  \end{array}\right)\,, \; \mathcal{D}(\vec{r}) =\left(  \begin{array}{cc}
    -2i \bar{\partial }& \alpha U(\vec{r}) \\ 
    \alpha U(-\vec{r}) & -2i \bar{\partial } \\ 
  \end{array}\right)\,, \label{mainmodel}
\end{align}

\no
where $\bar{\partial} = \frac{1}{2}(\partial_{x}+i\partial_{y})$ and  $U(\vec{r})=e^{-i\vec{q}_{1}\vec{r}}+e^{i\phi}e^{-i\vec{q}_{2}\vec{r}}+e^{-i\phi}e^{-i\vec{q}_{3}\vec{r}}$. The Hamiltonian $\mathcal{H}$ has only one dimensionless parameter $\alpha = w_{1}/(v_{0}k_{\theta})$ which fully controls the physics of the system. A similar idea of switching off the parameter $w_0$ was  investigated in Ref.\cite{Guinea2012}. It was argued there that  the Hamiltonian (\ref{mainmodel}) can be represented as $\mathcal{H}=\bm{\sigma}(-i\nabla +\alpha \vec{A})$ and viewed as  the  Hamiltonian for Dirac fermions propagating in  a background $SU(2)$ non-Abelian field $\vec{A}$.

 The band structure of the  Hamiltonian (\ref{mainmodel}) has some  remarkable properties. First, the Hamiltonian is particle-hole-symmetric, $\{\mathcal{H},\sigma_{z}\otimes 1\}=0$, which implies the spectrum to be symmetric with respect to zero energy. Second, the entire lowest 
band becomes absolutely flat ($\varepsilon_{0}(\vec{k})=0$ for all $\vec{k}$ in mBZ) at the recurrent values of  $\alpha$ corresponding to the magic angles $\theta$ of this model  (see Fig.~\ref{flatbands}). The first magic angle of this model is given by $\alpha_{1}\approx 0.586$, which corresponds to $\theta \approx 1.09^{\circ}$ on taking $w_{1}=110$meV and $2v_{0}k_{D}=19.81$eV. Moreover, the magic angle sequence in our model reveals a simple  quasiperiodicity in $\alpha$ with period  $\Delta \alpha \simeq 3/2$ (see Fig. \ref{magicperiod}). However, in the continuum model with  $w_{0} \ne 0$ this feature is smeared out with increasing  $w_0$ (see discussion below). All these remarkable 
features of the {\em chirally symmetric continuum model} (\ref{mainmodel}) indicate that this model captures the origin of the magic angles in the most precise way.

\begin{figure}
\includegraphics[width = 1.0 \columnwidth ]{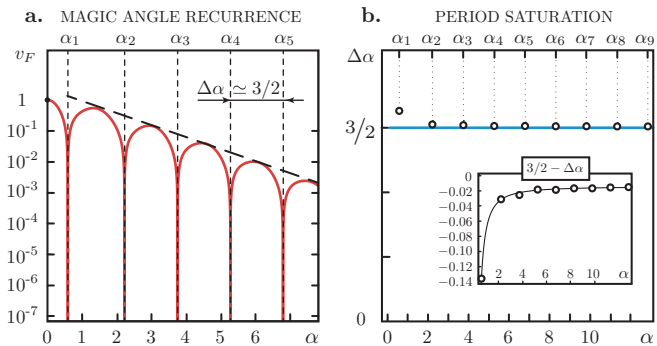}
\caption{Magic angle recurrence: a) Fermi velocity ($v_F=|\partial_{\vec{k}}E_{\vec{k}}|_{K,K'}$) at moire Dirac points as function of $\alpha$ (logarithmic scale). The first few magic angles are given by $\alpha = 0.586$, $2.221$, $3.751$,  $5.276$,
 $6.795$,  $8.313$, $9.829$, $11.345$, etc. We clearly see that the sequence of magic angles follows the approximate "3/2" rule: the distance between two adjacent $\alpha$ is "quantized" with approximately $3/2$ steps which saturate to $1.5$ as $\alpha \to \infty$, see  subfigure (b). } 
 \label{magicperiod}
\end{figure}

\textit{Zero mode equation and Fermi velocity.} We start from the fact that model (\ref{mainmodel})  has two zero modes at the points $K$ and $K'$ of the mBZ for arbitrary $\alpha$. This can be seen starting from  $ \alpha=0$. In this limit there are clearly four zero modes for eqn. \ref{mainmodel}, two each from the Dirac point in each layer. While the Dirac points in the two layers differ in crystal momentum, the pair of zero modes of each Dirac point differ in their  $C_3$  rotation eigenvalue $\omega=e^{i2\pi/3}$, and $\omega^* = e^{-i2\pi/3}$  see e.g. \cite{Po2018}. Thus each zero mode is uniquely labeled by symmetry eigenvalues. Furthermore, these symmetries commute with the particle-hole transformation $\sigma_z\otimes 1$. We can then consider each zero mode individually. On turning on $\alpha$ gradually, which preserves symmetry, each zero modes being unique must remain at zero energy. 

The equation for  the zero mode  at  the $K$ point,  $\mathcal{D}(\vec{r})\psi_{K}(\vec{r})=0$ reads in components 
\begin{align}
\label{zero mode 1}
\left(  \begin{array}{cc}
    -2i \bar{\partial }& \alpha U(\vec{r}) \\ 
    \alpha  U(-\vec{r}) & -2i \bar{\partial } \\ 
  \end{array}\right)\left(  \begin{array}{c}
    \psi_{K,1}(\vec{r}) \\ 
    \psi_{K,2}(\vec{r}) \\ 
  \end{array}\right) =0\,. 
  \end{align}
  
\no  
Obviously if $\psi_{K}(\vec{r}) = (\psi_{K,1},\psi_{K,2})^{\textrm{T}}$ is a solution of (\ref{zero mode 1}), then $\chi_{K}(\vec{r}) = \psi_{K}^{*}(-\vec{r}) $ is a solution to $\mathcal{D}^{*}(-
\vec{r})\chi_{K}(\vec{r})=0$. A general Bloch's wave function $\psi_{\vec{k}}(\vec{r})$ with  momentum $\vec{k}$ in mBZ has the form 
\begin{align}
\psi_{\vec{k}}(\vec{r}) = \sum_{m,n} \left(  \begin{array}{c}
    a_{mn} \\ 
    b_{mn} e^{i\vec{q}_{1}\vec{r}} \\ 
  \end{array}\right)e^{i(\vec{K}_{mn}+\vec{k})\vec{r}}\,, \label{Fourierform}
  \end{align}
where $\vec{K}_{mn} = m \vec{b}_{1}+n \vec{b}_{2}$ and $\vec{b}_{1,2} = \vec{q}_{2,3}-\vec{q}_{1}$ are the two Moir\'{e} reciprocal lattice vectors. The $K$ point corresponds to $\vec{k}=0$ and the renormalized Fermi velocity can be found through the first-order perturbation theory  
\begin{align}
v_{F}(\alpha) = \big|\partial_{\vec{k}} \frac{\langle \Phi_{K}|V_{\vec{k}}|\Phi_{K}\rangle }{\langle \Phi_{K}|\Phi_{K}\rangle}\big|_{\vec{k}=0}, \quad V_{\vec{k}} = \left(  \begin{array}{cc}
    0&  \bar{k} \\ 
   k & 0\\ 
  \end{array}\right)\,,
\end{align}
where $k,\bar{k}=(k_{x}\pm i k_{y})\sigma_{0}$ and $\Phi_{K}(\vec{r})=(\psi_{K},\chi_{K})^{\textrm{T}}$.  Using now relation  $\chi_{K}(\vec{r}) = \psi_{K}^{*}(-\vec{r})$, we find the expression for the Fermi velocity  as a function of $\alpha$,
\begin{align}
v_{F}(\alpha) =  \frac{|\langle \psi^{*}_{K}(-\vec{r})|\psi_{K}(\vec{r})\rangle| }{\langle \psi_{K}|\psi_{K}\rangle}\,. \label{Fermivel2}
\end{align}

\no
A striking result of this paper is however not just the vanishing of the Fermi velocity,  but the flattening of the entire lowest band. Below we show that it is possible to find the  absolute flat band 
solution because of a seemingly unrelated property, that the  entire zero mode spinor at the Dirac point, $\psi_{K}(\vec{r})$ vanishes exactly at the $BA$ stacking points and this happens precisely  at the magic angles.

\textit{Absolutely flat band.}  We now explain the origin of the absolutely flat band $\mathcal{H}\Phi_{\vec{k}}(\vec{r})=\varepsilon_{0}(\vec{k})\Phi_{\vec{k}}$, $\varepsilon_{0}(\vec{k})=0$  in our model. As follows from the Hamiltonian \eqref{mainmodel}, the appearance of the perfectly flat band at the set of magic angles 
implies that equation 
\begin{align}
\mathcal{D}(\vec{r})\psi_{\vec{k}}(\vec{r})=0 \label{flatbeq}
\end{align}

\no
has a solution for arbitrary Bloch's vector $\vec{k}$ in mBZ. As we explained above this equation always has  the zero mode solutions
$\psi_{K}(\vec{r})$ at the point $\vec{k}=0$. To relate solutions of  (\ref{flatbeq}) at arbitrary $\vec{k}$ to the zero-mode $\psi_{K}(\vec{r})$, we make a transformation to a new wave function $\eta_{\vec{k}}(\vec{r})=S(\vec{r})\psi_{\vec{k}}(\vec{r})$, with

\begin{align}
 S(\vec{r})=\frac{1}{\rho_{K}(\vec{r})} \left(  \begin{array}{cc}
    \psi_{K,2}(\vec{r})& -\psi_{K,1}(\vec{r})\\ 
    \psi^{*}_{K,1}(\vec{r}) &   \psi^{*}_{K,2}(\vec{r})\\ 
  \end{array}\right)\,,
\end{align}
and $\rho_{K}(\vec{r})=\psi^{\dag}_{K}\psi_{K}$ is the density of the zero mode wave function. Applying transformation $ S(\vec{r})$ to the operator $\mathcal{D}(\vec{r})$, one finds

\begin{align}
&S \mathcal{D}(\vec{r})S^{-1} =\left(  \begin{array}{cc}
    -2i\bar{\partial}-2i(\bar{\partial}\log \rho_{K})& 0\\ 
  h(\vec{r}) &   -2i\bar{\partial}\\ 
  \end{array}\right)\,, \label{maineqL}
\end{align}
where  $h(\vec{r})=\rho_{K}^{-1}(\vec{r})(\psi_{K,2}^{*})^{2}\big(2i\bar{\partial}(\psi_{K,1}^{*}/\psi_{K,2}^{*})+\alpha(U(-\vec{r})-U(\vec{r})(\psi_{K,1}^{*}/\psi_{K,2}^{*})^{2})\big)$. 
From equation (\ref{maineqL}) we see that the only possible solution for the first component is $\eta_{\vec{k},1}(\vec{r})=0$ \footnote{A general solution is
$\eta_{\vec{k},1}(\vec{r})=f(z)/\rho_{K}(\vec{r})$. Since the function $f(z)$ must obey Bloch-periodic boundary conditions it can be  non-trivial only if it has poles, but obviously $\rho_{K}(\vec{r}) \eta_{\vec{k},1}(\vec{r})$ is free of poles, so the only possible solution is  $f(z)=0$}. The latest  gives us an important relation between the flat-band wave function at the Dirac point $K$ and the flat-band wave function at an arbitrary mBZ point $\vec k$ precisely at the magic angles,
\begin{align}
\frac{\psi_{\vec{k},1}(\vec{r})}
{\psi_{\vec{k},{2}}(\vec{r})}= \frac{\psi_{K,1}(\vec{r})} {\psi_{K,{2}}(\vec{r})} \,.
\end{align}
Thus for the second component of the wave function  $\eta_{\vec{k}}(\vec{r})$  we have
\begin{align}
\bar{\partial}\eta_{\vec{k},2}(\vec{r}) =0\,, \label{maineqL2}
\end{align}
where $\eta_{\vec{k},2}(\vec{r})=
\psi_{\vec{k},1}(\vec{r})/\psi_{K,1}(\vec{r})=\psi_{\vec{k},2}(\vec{r})/\psi_{K,2}(\vec{r})$, which obeys
 the Bloch-periodic boundary conditions
 \begin{align}
 \eta_{\vec{k},2}(\vec{r}+\vec{a}_{1,2})=e^{i\vec{k}\vec{a}_{1,2}}\eta_{\vec{k},2}(\vec{r})\,, \label{boundcond}
 \end{align}

 \begin{figure}
\includegraphics[width = 0.8 \columnwidth ]{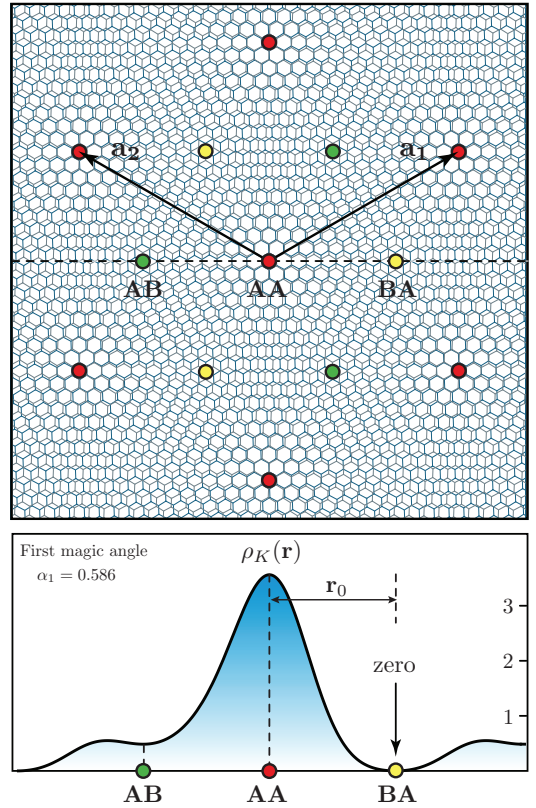}
\caption{(top) Schematic moire pattern with regions referred to in the text marked. (bottom) Wave function density $\rho_{K}(\vec{r})=\psi^{\dag}_{K}\psi^{}_{K}$ in real space for a single zero mode at the Dirac point:   $\rho_{K}(\vec{r})$ is localized on AA stacking and (exactly at the magic angles) has zeros on the BA stacking locations.  } 
 \label{magicperiod}
\end{figure}

\no 
where  $\vec{a}_{1,2}=\frac{4\pi}{3k_{\theta}}(\pm \frac{\sqrt{3}}{2},\frac{1}{2})$ are two Moir\'{e} lattice vectors. Equation (\ref{maineqL2}) may have a non-trivial solution if only the entire spinor $\psi_{K}(\vec{r})$ become zero at some point. We show below that  exactly at the angles where $v_{F}(\alpha)=0$, $\psi_{K,1}(\vec{r})$ and $\psi_{K,2}(\vec{r})$ do both  become zero at  the point  $\vec{r}_{0} = \frac{1}{3}(\vec{a}_{1}-\vec{a}_{2})$ which correspond to $BA$ stacking point (see Fig. 3).
Therefore we can find a meromorphic solution,
\begin{align}
\eta_{\vec{k},2}(\vec{r}) = \frac{\vartheta_{\frac{\vec{k}\vec{a}_{1}}{2\pi}-\frac{1}{6},\frac{1}{6}-\frac{\vec{k}\vec{a}_{2}}{2\pi}}(z/a_{1}|\omega)}{
\vartheta_{-\frac{1}{6},\frac{1}{6}}(z/a_{1}|\omega)}\,, \label{mainsol}
\end{align}

\no
 where $z=x+iy$, $a_{1}=(\vec{a}_{1})_{x}+i(\vec{a}_{1})_{y}$, $\omega =e^{i\phi}$ and $\vartheta_{a,b}(z|\tau)$ is the  theta function with rational characteristics $a$ and $b$ (see e.g. Ref.\cite{Mumford1983}),
  \begin{align}
\vartheta_{a,b}(z|\tau)=\sum_{n=-\infty}^{+\infty} e^{i\pi \tau(n+a)^{2}}e^{2\pi i(n+a)(z+b)}\,.
\end{align}

\no
Using the properties of the  theta function \cite{Mumford1983}, one can verify   that the meromorphic solution (\ref{mainsol})  obeys the periodic boundary conditions (\ref{boundcond}). 
Thus at the magic angles  the wave functions $\psi_{\vec{k}}(\vec{r})$ of the exactly flat  band read

 \begin{align}
\psi_{\vec{k}}(\vec{r}) =  \frac{\vartheta_{\frac{\vec{k}\vec{a}_{1}}{2\pi}-\frac{1}{6},\frac{1}{6}-\frac{\vec{k}\vec{a}_{2}}{2\pi}}(z/a_{1}|\omega)}{
\vartheta_{-\frac{1}{6},\frac{1}{6}}(z/a_{1}|\omega)}\psi_{K}(\vec{r})\,. \label{mBzflatband}
\end{align}

\no
Note that under this construction, the zeros of $\psi_{K}(\vec{r})$ \textit{exactly} cancel  zeros of the theta function in the denominator. 

Therefore, exactly at the magic angles, where $\psi_{K}(\vec{r_0})=0$, the wave functions \eqref{mBzflatband} satisfy  the zero-mode equation \eqref{flatbeq} for all $\vec k$ in mBZ. Thus we showed that there is an perfectly flat band solution, $\varepsilon_0(\vec k) \equiv 0$.

\textit{Connection with the vanishing of Fermi velocity. } Now we  show that zero  Fermi velocity is connected to  zero of $\psi_{K}(\vec{r})$.  Analyzing symmetries of the zero-mode operator $\mathcal{D}(\vec{r})$ one can check that if $\psi_{K}(\vec{r})$ is a solution to the equation $\mathcal{D}(\vec{r})\psi_{K}(\vec{r})=0$, then
$\psi_{K}(R_{\phi}\vec{r})$ is also a solution, where $R_{\phi}$ denotes a counterclockwise rotation on angle $\phi$. Similarly one finds that 
 \begin{align}
\mathcal{D}(\vec{r}\pm\vec{r}_{0})\left(   \begin{array}{c}
    \psi_{K,1}(R_{\phi}\vec{r}\pm\vec{r}_{0}) \\ 
    e^{\mp i\phi} \psi_{K,2}(R_{\phi}\vec{r}\pm\vec{r}_{0}) \\ 
  \end{array}\right)=0\,,
\end{align}
where $\vec{r}_{0}= \frac{1}{3}(\vec{a}_{1}-\vec{a}_{2})$. Since at  $\alpha=0$ we have $\psi_{K}(\vec{r})=(1,0)$ this implies for the zero-mode components 
at  arbitrary $\alpha$
 \begin{align}
&\psi_{K,1}(R_{\phi}\vec{r}\pm\vec{r}_{0})=\psi_{K,1}(\vec{r}\pm\vec{r}_{0})\,, \\
&\psi_{K,2}(R_{\phi}\vec{r}\pm\vec{r}_{0})=e^{\pm i\phi}\psi_{K,2}(\vec{r}\pm\vec{r}_{0})\,.
\end{align}
The second equation implies that $\psi_{K,2}(\vec{r})=0$ at $\vec{r}=\pm \vec{r}_{0}$ for arbitrary $\alpha$. Now to relate appearance of zeros in $\psi_{K}(\vec{r})$ to  zeros of the renormalized Fermi velocity, we notice that the Fermi velocity is proportional to an integral of motion of the operator $\mathcal{D}(\vec{r})$\footnote{One can derive it by multiplying  equations in (\ref{zero mode 1}) by $\psi_{K,1}(-\vec{r})$ and $\vec{\psi}_{K,2}(-\vec{r})$ and adding to them their reversed counterparts. }
\begin{align}
v(\alpha) = \psi_{K,1}(\vec{r})\psi_{K,1}(-\vec{r})+\psi_{K,2}(\vec{r})\psi_{K,2}(-\vec{r})\,,
\end{align}
where $v(\alpha)$ does not depend on coordinates and from (\ref{Fermivel2}) we see that $v_{F}(\alpha)\sim v(\alpha)$.  Using that $\psi_{K,2}(\pm\vec{r}_{0})=0$ we find 
 \begin{align}
v_{F}(\alpha) \sim  \psi_{K,1}(\vec{r}_{0})\psi_{K,1}(-\vec{r}_{0})\,.
\end{align}

\no
And one can see from the equations of motion near the point $\vec{r}=-\vec{r}_{0}$ that $\psi_{K,1}(-\vec{r}_{0})$ can not be zero. Therefore $v_{F}(\alpha)=0$ means that  $\psi_{K,1}(\vec{r}_{0})=0$ and vice versa. Thus we finally obtain that $v_{F}(\alpha)=0$ implies the existence of an absolutely flat band, whose wave functions are given by (\ref{mBzflatband}).

The appearance of zeros in $v_{F}(\alpha)$ is not surprising, since this is just a real function of a single parameter. By varying this parameter 
one hopes that $v_{F}(\alpha)$ crosses zero at some value of $\alpha$. To check analytically that this actually happens in our model near $\alpha_{1}\approx 0.586$
we use perturbation theory in $\alpha$.

\

\textit{Perturbation theory in $\alpha$. $K$ point.}  One can analyze the zero mode equation (\ref{zero mode 1}) using perturbation theory in $\alpha$, namely the spinor $\psi_{K}(\vec{r})$ should have the form
\begin{align}
\psi_{K}(\vec{r})=\left(  \begin{array}{c}
    \psi_{K,1} \\ 
    \psi_{K,2} \\ 
  \end{array}\right) = \left(  \begin{array}{c}
    1+\alpha^{2}u_{2}+\alpha^{4}u_{4}+\dots \\ 
    \alpha u_{1}+\alpha^{3} u_{3}+\dots \\ 
  \end{array}\right) \,.
\end{align}
In general we can find $u_{n}(\vec{r})$ step by step to arbitrary  order in $\alpha$. Limiting ourselves to the first lowest terms we find 
 \begin{align}
  &u_{1} =  -i  (e^{i\vec{q}_{1}\vec{r}}+e^{i\vec{q}_{2}\vec{r}}+e^{i\vec{q}_{3}\vec{r}})\,, \notag\\
 &u_{2} = \frac{i}{\sqrt{3}}e^{-i\phi}(e^{-i\vec{b}_{1}\vec{r}}+e^{i\vec{b}_{2}\vec{r}}+e^{i(\vec{b}_{1}-\vec{b}_{2})\vec{r}})+c.c.
\end{align}

\no 
And   perturbation formula for the renormalized Fermi velocity reads
\begin{align}
v_{F}(\alpha) =\frac{1-3 \alpha^{2}+\alpha ^4-\frac{111 \alpha ^6}{49}+\frac{143 \alpha ^8}{294}+\dots}{1+3 \alpha^2+2 \alpha ^4+\frac{6 \alpha ^6}{7}+\frac{107 \alpha ^8}{98}+\dots}\,.
\end{align}
This expression gives for  the first magic angle $\alpha_1 \approx 0.587$, which is very close to the precise value $\alpha_{1}\approx 0.586$.  
Therefore the perturbation theory for small $\alpha$ quantitively explaining the appearance of $\alpha_1$. Note that up to $\alpha^2$, $v_{F} \approx (1 -3 \alpha^2)/(1+6 \alpha^2) $ was reported in a model with $w_{0}=w_{1}$  \cite{Bistritzer2010}.

\textit{$\Gamma$ point.} In our notations the $\Gamma$ point corresponds to $\vec{k}=\vec{q}_{1}$ in (\ref{Fourierform}). The symmetries of the Hamiltonian (\ref{mainmodel}) imply that $\chi_{\Gamma}(\vec{r}) = \lambda_{\alpha}\sigma_{x}\psi_{\Gamma}(\vec{r})$, where $\lambda_{\alpha}=\pm 1$, moreover one can also obtain that $\psi_{\Gamma,2}(\vec{r}) = i \mu_{\alpha}\psi_{\Gamma,1}(-\vec{r})$, where $\mu_{\alpha}=\pm 1$. Using the last equality and the equation (\ref{Fermivel2}) one immediately finds that the velocity at the $\Gamma$ point is exactly zero, which means that the $\Gamma$ point is the extremum for all bands.
Finally using all symmetry relations we find that the whole spectrum at the $\Gamma$ point is characterized by two equations
\begin{align}
 \quad 2\bar{\partial}\psi_{\Gamma,1} \mp \alpha U(\vec{r})\psi_{\Gamma, 1}(-\vec{r}) = \varepsilon_{\Gamma} \psi_{\Gamma,1}(-x,y)\,,\label{Gammaeq}
\end{align}

\no
where the "$-$" captures all odd magic angles and "$+$" all even.   
One can study  the equations (\ref{Gammaeq}) perturbatively for small $\alpha$ and find for the energy at the $\Gamma$ point  (which equals to half-bandwidth \footnote{We find numerically that in the model with $w_{0}=0$ the energy at the $\Gamma$ point is a global maximum of the lowest band.})

\begin{figure}
\includegraphics[width = 0.9 \columnwidth ]{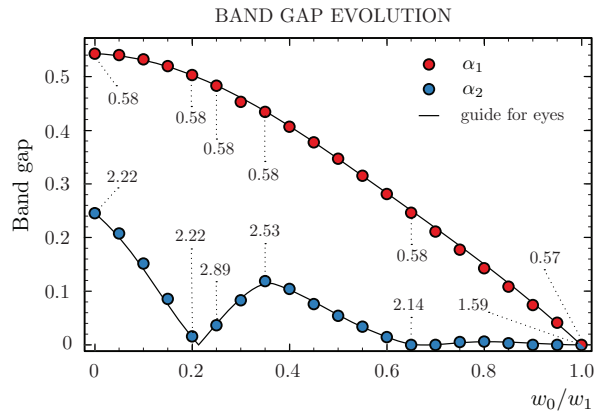}
\caption{Band gap evolution in the flow from CS-CM  ($w_0=0$) to BM-CM ($w_0=w_{1}$) model. The first magic angle (red dots) remains stable throughout all the region, $\alpha_1 \approx 0.58$, while both the second magic angle (blue dots) and the corresponding band gap experiences oscillations and discontinuities. The numbers with the dotted lined signify the local values of an approximate magic angle for the give $w_0/w_1$. }
\label{gap_evolution}
\end{figure}

\begin{align}
\varepsilon_{\Gamma} = 1-2\alpha +\frac{\alpha^{2}}{3}+\frac{2 \alpha ^3}{9}+\frac{5 \alpha ^4}{54}+\dots\,, \label{spGamma}
\end{align}
and the wave function 
\begin{align}
\psi_{\Gamma,1}(\vec{r})& = U(-\vec{r}) +\frac{\alpha}{3}U(2\vec{r})+\frac{\alpha^{2}}{18}\big((2-e^{i\phi})U(-\sqrt{7}R_{\gamma}\vec{r})\notag\\
&+(2-e^{-i\phi})U(-\sqrt{7}R_{-\gamma}\vec{r})-4U(2\vec{r})\big)+ \dots\,,
\end{align}
where $R_{\gamma}\vec{r}$ is a counterclockwise rotation on angle $\gamma$ with $\tan(\gamma) =\sqrt{3}/5$.
We see that the equation (\ref{spGamma}) gives $\alpha_{1}\approx 0.585$.

\textit{Turning on $w_{0}$}. We  turn on the $w_{0}$ terms in the $T(\vec{r})$ matrix (\ref{Tj}), but still neglect relative rotations in the kinetic terms $\bm{\sigma}_{\pm \theta/2}\to \bm{\sigma}$. Then the dimensionless Hamiltonian  has the form 
\begin{align}
\mathcal{H}_{w_{0}} = \mathcal{H} +(w_{0}/w_{1}) \sigma_{0}\otimes V(\vec{r})\,,
\end{align}
where 
\begin{align}
 V(\vec{r})= \left(  \begin{array}{cc}
   0& U_{0}(\vec{r})\\ 
    U^{*}_{0}(\vec{r}) &  0\\ 
  \end{array}\right)\,,
\end{align}
and 
$U_{0}(\vec{r})=e^{-i\vec{q}_{1}\vec{r}}+e^{-i\vec{q}_{2}\vec{r}}+e^{-i\vec{q}_{3}\vec{r}}$. 
We present numerical dependence for the gap between the lowest band and the next excited band at the first two magic angles as a function of $w_{0}/w_{1}$ (see Fig. \ref{gap_evolution}). Importantly, for the first magic angle the band gap vanishes at $w_{0}=w_{1}$.

\textit{Relevance to Experiments.} We briefly discuss the importance of our model for experiments. First,  we notice analytically and verify numerically that the first magic angle remains almost unchanged ($\alpha_1 \approx 0.586$) even if $w_0$ is continuously tuned on to approach the BM-CM models. In CS-CM model, however, the value of the first magic angle is $1.09^{\circ}$ vs $1.05^{\circ}$ in BM-CM. The former, perhaps fortuitously, is closer to the value reported in experiments \cite{Cao2018a,Cao2018b}. It is important to make connection with Ref.\cite{Koshino2018}, where it was argued that lattice relaxation effects shrinks the AA regions, thus reducing the value of $w_0$  by approximately 20\%. Also we note, that due to finite and relatively large gap to the higher bands observed experimentally \cite{Cao2018a,Cao2018b} qualitatively resembles the CS-CM model.   Furthermore, we predict the second magic angle (the angle where both the bandwidth has minimum and the band gap has maximum), occurs around $\alpha_2 \approx 2$, which converts to $\theta \approx 0.22^\circ$-$0.29^{\circ}$, depending on the precise value of the ratio $w_0/w_1$. It would be interesting to search for this angle in future experiments.

\

\textit{Discussion. - } In this paper, we introduced a variant of the continuum model used to describe TBG, and show that the notion of magic angles acquires a remarkably robust character visible in several properties including the perfect flatness of the band at neutrality. 
We showed that the existence of the flat band is related to the appearance of zeros in the real space Bloch wave functions at the moire $K$ (Dirac) point.  Given that the model has flat bands, the appearance of the first magic angle can be precisely captured in perturbation theory in $\alpha$, i.e. we find $\alpha$ when the bandwidth goes to zero.  
With chiral symmetry, the single valley band structure is known to be anomalous \cite{Po2018} (surface state of a three dimensional topological insulator in class AIII) which is only permitted since we begin with a continuum rather than a lattice model. The role of this anomaly in the flat band phenomena would be of great interest to explore. 
 Our minimalistic model lies within a general class of continuum  models for  TBG allowed by symmetries. Other models (such as e.g. \cite{Santos2007,Bistritzer2010}) can be viewed as including perturbation on top of our model.  The flat band model captures the essence of the flatness phenomena and thus the source of magic angles in a convenient way, and points to a rich underlying mathematical structure, uncovering which is left to future work.

\

\textit{Acknowledgments. - } We thank Bertrand Halperin and  Subir Sachdev for useful comments.   A. V. would like to thank Adrian Po, Liujun Zhou, T. Senthil and YiZhuang You for several discussions on TBG. G.T.  was  supported  by the MURI grant W911NF-14-1-0003 from ARO and by DOE grant de-sc0007870. A.J.K. was supported by the Swiss National Science Foundation's grant  P2ELP2\_175278.  A.V. was supported by a Simons Investigator award and by NSF-DMR 1411343.

\bibliographystyle{ieeetr}
\bibliography{Refs}

\end{document}